\documentclass[sigconf]{acmart}
\acmConference[SE 2030]{International Workshop on Software Engineering in 2030}{November 2024}{Puerto Galinàs (Brazil)}

\usepackage{graphicx} 
\usepackage{listings}
\usepackage{pifont}
\usepackage{todonotes}

\title{Test Oracle Automation in the era of LLMs}

\author{Facundo Molina}
\affiliation{
  \institution{IMDEA Software Institute}
  \city{Madrid}
  \country{Spain}
}

\author{Alessandra Gorla}
\affiliation{
  \institution{IMDEA Software Institute}
  \city{Madrid}
  \country{Spain}
}

\date{February 2024}

\definecolor{dkgreen}{rgb}{0,0.6,0}
\definecolor{gray}{rgb}{0.5,0.5,0.5}
\definecolor{mauve}{rgb}{0.58,0,0.82}

\definecolor{mygray}{gray}{1}

\begin{document}

\begin{abstract}

The effectiveness of a test suite in detecting faults highly depends on the correctness and completeness of its test oracles.
Large Language Models (LLMs) have already demonstrated remarkable proficiency in tackling diverse software testing tasks, such as automated test generation and program repair.
This paper aims to enable discussions on the potential of using LLMs for 
test oracle automation, along with the challenges that may emerge during the generation of various types of oracles.
Additionally, our aim is to initiate discussions on the primary threats that SE researchers must consider when employing LLMs for oracle automation, encompassing concerns regarding oracle deficiencies and 
data leakages.
\end{abstract}

\maketitle

\section{Introduction}

The goal of software testing is to find defects. 
To actually find defects, though, test suites require relevant test inputs, i.e., 
inputs that can exercise the software under test (SUT) 
in realistic scenarios. Furthermore, to ensure that 
the SUT exhibits the expected behavior for these inputs,
accurate test oracles are required. The problem 
of automating the generation of test oracles, the so-called
oracle problem~\cite{Barr+2015}, has become very relevant
in the last decades, as it has the potential to improve 
the oracles used in the testing process, and therefore
contributing to revealing defects in software.

Various approaches have been proposed to address the oracle problem by 
automatically deriving different kinds of oracles~\cite{DBLP:journals/scp/ErnstPGMPTX07,autoinfer2011,gassert2020,evospex2021,memo2021,precis2021,toga2022, specfuzzer2022, agora2023}, including 
\emph{test assertions}, \emph{contracts} (such as pre/postconditions and invariants) or 
\emph{metamorphic relations}. 
Generally, these approaches 
observe some artifact related to the SUT (documentation, comments, 
source code, executions) and then derive oracles that 
are consistent with the observations. For example, 
TOGA~\cite{toga2022} observes the source code 
of a target test and a focal method (method under test)
and infers a test assertion for the given test; 
MeMo~\cite{memo2021} extracts metamorphic relations by observing natural language comments in the source code; 
Daikon~\cite{DBLP:journals/scp/ErnstPGMPTX07} and related tools \cite{gassert2020,evospex2021,specfuzzer2022} observe the behavior of the SUT 
(from a set of tests) in order to infer class invariants and pre/postconditions.

Despite all these efforts, the problem of automatically deriving oracles is still
an open research problem
in software testing~\cite{Barr+2015}.
The main reason is that automatically derived oracles are rarely accurate.
These accuracy issues can result in high false positive rates, 
which can lead to false alarms and reduce the trustworthiness of the testing process. 
Indeed, recent studies have shown that even state-of-the-art neural based approaches can produce oracles with high false positive rates~\cite{DBLP:conf/sigsoft/HossainFDEV23}.


Large Language Models (LLMs) have already been used
in tackling diverse software testing~\cite{testing-llms-2024}, 
demonstrating remarkable proficiency, particularly in tasks
such as automated program repair~\cite{DBLP:conf/icse/FanGMRT23, DBLP:conf/icse/XiaWZ23, DBLP:conf/icse/JiangLLT23}
and automated test generation~\cite{cat-lm-2023, codamosa2023, DBLP:journals/tse/SchaferNET24}. 
Given the positive results, researchers have also started to explore 
the use of LLMs to automatically generate test oracles, 
mainly in the form of test assertions~\cite{DBLP:conf/ast/TufanoDSS22, teco2023,DBLP:conf/icse/NashidSM23}. 
Although the initial results are promising, showing that generated 
assertions can improve test coverage~\cite{DBLP:conf/ast/TufanoDSS22} and 
some lexical/functional metrics~\cite{teco2023}, there are still aspects of 
LLM-generated oracles that need to be further explored.

This paper aims to enable discussions on the use of LLMs for 
test oracle automation in general (not only test assertions) 
by discussing:
\begin{itemize}
  \item the potential of LLMs for test oracle automation (including oracles that go beyond test assertions, such as contracts or metamorphic relations) and the challenges when using the LLMs through prompt engineering or by pre-training or fine-tuning them; and
  \item the main threats that arise from the use of LLMs to generate different 
  kinds of oracles, including oracle deficiencies and privacy-related issues related to data leakages, 
  and how we can mitigate such threats.
\end{itemize} 


\section{LLMs for Oracle Automation}


The most straightforward application of LLMs for oracle automation involves prompt engineering, wherein a prompt is designed to instruct any state-of-the-art pre-trained model, such as ChatGPT-3.5~\cite{openai} or Llama2~\cite{llama2}, to produce an oracle. 
The majority of studies in the literature that utilize LLMs for software testing tasks, as well as for other applications in general, concentrate on two primary strategies for prompt engineering: zero-shot learning and few-shot learning~\cite{testing-llms-2024}.
Zero-shot learning consists of providing a prompt 
asking the model for results, e.g., 
``Generate assertions for the following test: ...''.
Few-shot learning, instead, consists of providing a set 
of high-quality examples to the model. For instance, one could provide a set of test-assertion pairs for the model to generate assertions for a given test. 

A more sophisticated approach to oracle automation via LLMs involves \textbf{pre-training or fine-tuning}. 
Pre-training entails training the model on a broad distribution of data to predict the subsequent token in a sequence. Conversely, during fine-tuning, the weights of a pre-trained model are adjusted by retraining it on a designated dataset tailored for a specific task. 
A clear example of this approach has been proposed by Tufano et al.~\cite{DBLP:conf/ast/TufanoDSS22}, where an LLM is 
pre-trained with a large source code and English language corpora, 
and then fine-tuned for generating assert statements.

Regardless of the strategy used, 
generating oracle via LLMs requires 
the use of data related to the expected oracles, 
either to build prompts or to pre-train or fine-tune the model. 
Various sources of pertinent information (such as source code, test code, documentation, logs, etc.) may be utilized to feed the LLMs. 
However, the type of information employed will also vary based on the type of oracle intended to be generated, thereby presenting distinct challenges. For instance, employing a zero-shot learning approach to produce a test assertion for a given test merely requires furnishing a prompt containing the test case and instructing the model to expand it with assertions.
Applying the same methodology to generate broader oracles, such as postconditions for a method, necessitates a more detailed prompt. In addition to incorporating the method code, one must also specify the formalism or language for articulating the contract (e.g., as a code fragment, an assert statement, a logical expression, etc.).

To properly illustrate these challenges,
we now consider the most common types of 
automatically inferred oracles:
test assertions, contracts, and metamorphic relations.

\subsection{Test Assertions}

Test assertions in unit tests check 
the expected behavior of the SUT in a specific scenario, 
and are typically expressed as code. 
Given their importance for having high quality test suites, 
the automated generation of test assertions has been widely studied, 
including approaches that pre-train and fine-tune the models~\cite{DBLP:conf/ast/TufanoDSS22, teco2023}. 
Tufano et al.~\cite{DBLP:conf/ast/TufanoDSS22} pre-trained a BART 
Transformer model~\cite{DBLP:journals/corr/abs-1910-13461} 
with a large corpus of English text and Java code, 
and then fine-tuned it on the task 
of generating assert statements for unit test cases. 
Similarly, TeCo~\cite{teco2023} fine-tunes 
the CodeT5~\cite{DBLP:conf/emnlp/0034WJH21} and CodeGPT~\cite{lu2021codexglue} 
LLMs specifically for the test completion task
(i.e., predict the next statement in a test case), 
which are then evaluated for assert statement generation 
from the code under test (including the method under test), 
the test method signature, and the prior statements before the
assertion statement.

These approaches can achieve an exact match rate 
(percentage of generated assertions that exactly match the 
expected assertions) of up to 62\%.
The recent improvements achieved by state-of-the-art LLMs enable 
the use of such models for generating test assertions using 
a more straightforward approach, such as zero-shot learning.
Consider for example the test in Figure~\ref{stack-test}, 
which shows a very simple test case for a \texttt{Stack} class, 
with a unique test assertion checking that the stack is not empty after
three push operations and one pop operation.

\begin{figure}[h]
\begin{lstlisting}[]
public void testPop() {
  Stack<Integer> stack = new Stack<>();
  stack.push(2);
  stack.push(3);
  stack.push(5);
  stack.pop();
  assertFalse(stack.isEmpty());
}
\end{lstlisting}
\caption{A simple test for a \texttt{Stack} class.}
\label{stack-test}
\end{figure}

If we provide the prompt 
``Extend the following Java test just with assert statements: + test-code''
where test-code is the test \texttt{testPop}, 
ChatGPT-3.5 produces the following assertions:
\newline
\begin{center}
\begin{lstlisting}[frame=none]
assert !stack.isEmpty() : "Stack should not be empty after pop";
assert poppedElement == 5 : "Popped element should be 5";
\end{lstlisting}
\end{center}
\vspace{0.1cm}
also checking that the popped element is 5
(ChatGPT-3.5 edited the test to save
the result of pop in the variable \texttt{poppedElement}). 
In fact, Nashid et al~\cite{DBLP:conf/icse/NashidSM23} recently 
proposed a technique for prompt creation based
on embedding or frequency analysis, that can achieve  
an exact match rate of 76\% in test assertion generation.

\subsection{Contracts} 

Contracts~\cite{DBLP:journals/computer/Meyer92} are logical constraints on a specific
software element (method, class, etc.), 
and are usually captured in the form of preconditions 
and postconditions for methods, or representation invariants for classes.
As opposed to test assertions, contracts are not specific to a particular 
test case, but rather express properties that must hold for any execution.
Moreover, contracts can be expressed in different formalisms, 
not only in the same language as the SUT.
For instance, JML~\cite{DBLP:conf/fmco/ChalinKLP05} is a 
popular library for Java that includes a language for writing 
contracts as annotations in the source code. 
Daikon~\cite{DBLP:journals/scp/ErnstPGMPTX07}, 
a well-known tool for dynamic invariant detection, 
produces pre/postconditions and class invariants 
using its own language (a mix of Java, 
and mathematical logic). 
Other contract inference tools also use their own 
languages~\cite{gassert2020,evospex2021,specfuzzer2022}.

To the best of our knowledge, there are still no studies 
using LLMs to generate contracts.
To give an idea on how LLMs could be used for this task,
let us consider an implementation of the \texttt{push}
operation for a \texttt{Stack} class, 
shown in Figure~\ref{stack-push}. 

\begin{figure}[h]
\begin{lstlisting}[]
public void push(E e) {
  if (size == elements.length) {
    ensureCapacity();
  }
  elements[size++] = e;
}
\end{lstlisting}
\caption{Implementation of a \texttt{push} operation for a \texttt{Stack}.}
\label{stack-push}
\end{figure}

The code first ensures that there is enough capacity 
in the \texttt{elements} array, and then pushes the element 
in the next available position. A contract for this method
could include a postcondition stating that the size of the stack
is increased by one, and that the element is added at the top (\texttt{size-1}) of the stack.

The variety of formalisms for expressing contracts poses a challenge 
when using LLMs, 
mainly because one would also need to specify the formalism
in which the contract must be expressed. 
State-of-the-art pre-trained models have the potential 
of producing contracts with a reasonable performance, 
using a prompt engineering approach,  
at least for well-established formalisms. 
For instance, with the zero-shot prompt 
``Generate a postcondition in the form of an assert statement for the following method: + method-code'',
where method-code is the \texttt{push} method in Figure~\ref{stack-push}, 
ChatGPT-3.5 produces the following postcondition:
\newline
\begin{lstlisting}[frame=none]
assert elements[size - 1].equals(e) : "Element was not successfully pushed onto the stack.";
\end{lstlisting}
\vspace{0.3cm}



Note that although the postcondition captures one expected property, 
its execution will crash if the pushed element is \texttt{null}. 
Similarly, if we ask for the postcondition to be expressed in JML,
the model produces the same postcondition. 

However, for less known formalisms or subjects with a more complex API, 
using a few-shot prompt, 
or even pre-training and fine-tuning the model, 
could be more appropriate.
While using a few-shot prompt would 
require one to provide a set of examples of the expected contracts
(e.g., a set of methods with their contracts expressed in the desired formalism),
the use of pre-training or fine-tuning would require
one to provide a specific and extensive dataset for contract inference,
including the software element (method or class), the expected contract,
and possibly other contextual information related to the contract 
to learn.



\subsection{Metamorphic Relations}

Metamorphic relations express 
domain-specific properties 
of multiple executions of the SUT~\cite{Segura2016}.
Compared to test assertions and contracts,
they are more general oracles that 
are easy to define and mantain.
Some examples of metamorphic relations involving two executions are 
$p(x,y) = p(y,x)$ for a commutative operation $p$, or
$sort(a) = sort(b)$ for a sorting operation where $a$
is a permutation of $b$.
Since its introduction by Chen et al.~\cite{chen1998},
metamorphic relations
have been successfully used for detecting bugs
in a variety of software systems, 
including the search engines Google and Bing~\cite{DBLP:journals/tse/ZhouXC16}
and the Web APIs of Spotify and YouTube~\cite{DBLP:journals/tse/SeguraPTC18}.
Moreover, metamorphic relations have shown to be
complementary to other types of oracles, such as test assertions, 
to detect faults in the SUT~\cite{memo2021}.

According to the literature~\cite{Segura2016}, 
metamorphic relations are expressed through 
some formalism that allows to capture the expected 
relationship between the inputs and outputs. 
Metamorphic relations are typically instantiated in a test case, 
where a \emph{source test} (e.g., $a1 = sort([1,3,2])$)
is executed, then a \emph{follow-up test} (e.g., $a2 = sort([2,1,3])$) is executed,
and finally the relation is checked ($a1 = a2$)~\cite{segura2020}.

For the inference of metamorphic oracles, 
the use of LLMs has not yet been explored.
Using LLMs to generate metamorphic relations 
can be challenging, 
mainly because of the domain-specific knowledge required to identify the relations, 
and the lack of well-defined formalisms to express them.
A straightforward approach could be enabled in the case that the 
SUT already contains a set of test cases, as these 
could be used as source tests to ask  
an LLM to generate follow-up tests that preserve
some relation with respect to them.
For example, using the zero-shot prompt 
``Generate a follow-up test that is equivalent to the following test + test-code''
with the \texttt{testPop} test from Figure~\ref{stack-test},
ChatGPT-3.5 produces a test with exactly the same operations,
and two additional sentences:
\newline
\begin{lstlisting}[frame=none]
stack.push(7);
stack.pop();
\end{lstlisting}
\vspace{0.1cm}
The produced follow-up test is equivalent to the source test in the sense that
both result in the same stack. 
It is easy to see that
the metamorphic relation behind states that
for every stack, if we push an element and then pop it, the stack remains the same.
Using both tests, one could easily implement a new test to check the metamorphic relation
by first executing the source test, then the follow-up test, and finally checking that the stacks
are equal.

If a set of test cases is not available, 
or one wants to focus on generating the metamorphic 
relation itself (instead of its implementation), 
one would need to include in the communication with the LLM how the relations should be expressed. 
For this scenario, a few-shot prompt or a 
pre-training and fine-tuning approach would be necessary. 
Independently of the kind of oracle we are generating,
it is evident that LLMs have an enormous potential to
assist in oracle automation.
However, as we discuss in the next section,
the use of LLMs for this task not only inherits some 
threats from previous techniques 
that can affect the quality of the produced oracles, 
but also introduce new threats that need to be considered.

\section{Threats to Validity}

\begin{figure*}[t]
  \begin{center}
  \includegraphics[width=.80\textwidth]{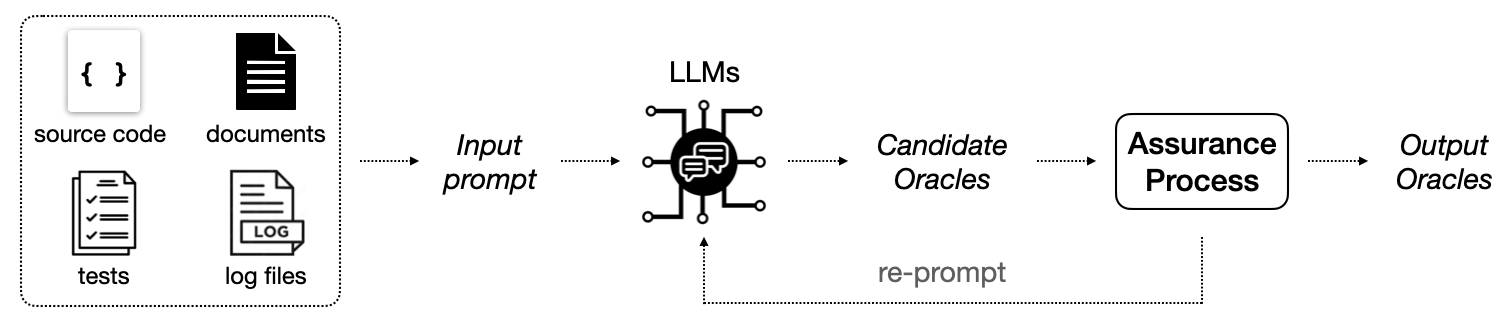}
  \end{center}
  \caption{Overview of LLM-based Oracle Generation.}
  \label{llm-oracle-gen}
\end{figure*}

\subsection{Oracle Deficiencies}

The quality of an oracle can be assessed in terms of  
its oracle deficiencies: false positives and false negatives~\cite{DBLP:conf/issta/JahangirovaCHT16}.
Intuitively, a \textbf{false positive} is a correct and expected program state
for which the oracle is false, i.e., a false alarm.
A \textbf{false negative} is an incorrect and unexpected program state
for which the oracle is true, i.e., a missed fault.
While false negatives are more tolerable, as one can
still fix the oracle to improve its fault detection capability, 
false positives are more critical, as they can lead to 
false alarms resulting in unnecessary debugging efforts. 

Failing to identify and eliminate false positives may result 
in high false positive rates, and raise concerns about  
the practical usefulness of the tools \cite{DBLP:conf/sigsoft/HossainFDEV23}.
To mitigate this, we consider that assurance mechanisms post-processing 
LLM-produced oracles are imperative to provide 
guarantees on their quality, 
and may be crucial to minimize oracle deficiencies.
Alshahwan et al.~\cite{Alshahwan2024} recently 
proposed the notion of Assured LLM-based Software Engeering 
(Assured LLMSE) with the aim of providing guarantees on 
the output produced by LLMs used for software engineering tasks.
Figure~\ref{llm-oracle-gen} shows an overview of
how Assured LLMSE could be applied 
to LLM-based oracle generation, 
where the LLM-produced oracles are subjected to an assurance process, 
possibly analyzing deficiencies. 
The assurance process can also provide information to re-prompt the model 
with those oracles that do not pass the process.

As LLMs offer absolutely no guarantees on their outcome, 
the assurance process could first easily eliminate 
syntactically incorrect oracles.
Detecting oracle deficiencies, however,
is more challenging.
False positives detection depends on 
the usage context.
If the oracles are for regression testing, i.e., 
to equip the current version of the SUT with oracles
in order to detect regression errors in the future,
to detect false positives one can search for a 
reachable program state (test case)
in which the oracle fails. 
For example, GAssert~\cite{gassert2020} uses evolutionary 
computation to actively search for test cases that 
falsify a given assertion oracle.
On the other hand, if the oracles are intended 
to test the current implementation, the detection of
false positives may 
require a human to decide whether an oracle failure 
indicates a false positive or a real bug.

Once false positives are detected, 
one can fix the oracle either by removing the parts 
that are incorrect, or by weakening the
oracle to make it more general. This could enable 
an iterative process, possibly re-prompting the LLM, 
in which the oracle is refined 
until no more false positives are detected.
This iterative refinement process has already been 
applied for postcondition assertions~\cite{gassert2020}. 

Although false negatives are less critical, 
their detection can help to 
strengthen the oracle,
improving its fault detection capability.
As false negatives are essentially faulty program states 
missed by the oracle, 
mutation analysis can be used to introduce artificial 
faults, and then check if the oracle can detect them.
While some approaches use mutation analysis to improve 
the oracles during the inference process~\cite{DBLP:conf/issta/FraserZ10, gassert2020,specfuzzer2022},
others use it as an evaluation technique
for fault detection analysis~\cite{memo2021, agora2023}. 
Similarly, LLM-based techniques could also benefit from false 
negatives analysis.

\subsection{Oracle Leakages}

The use of LLMs for software engineering tasks 
has several threats, 
including closed-source models, 
data leakages between training data and research evaluation,
and reproducibility issues~\cite{Sallou2023}.
Among these, data leakages is particularly 
relevant for LLM-based oracle automation, 
specifically during the evaluation of these approaches.

As LLMs are trained on very large amounts of data,
often including publicly available code from GitHub~\cite{chen2021evaluating}, 
there is a risk that the model has memorized some 
code samples from the training data~\cite{inan2021training}. 
Thus, when evaluating LLM-based techniques for oracle automation,
we need to pay special attention to the data we use 
for evaluation, as we may end up obtaining oracles 
that are not actually created by the model,
but rather replicated from the training data.

To illustrate this issue, let us consider Defects4J~\cite{defects4j}, one the 
most widely-adopted benchmarks in 
software testing research.
Many of the projects involved in Defects4J are 
publicly available on GitHub. Thus, evaluating LLM-based
oracle generation techniques on Defects4J 
can clearly lead to oracle leakages, and make the LLM 
provide accurate oracles just because they are 
a copy of the oracles from the training data. 
Figure~\ref{arraystack-collections} shows an example of 
a test prefix (i.e., the first part of a test without the
assertions) from the Apache Commons Collections 
project in Defects4J, available in revision 
\texttt{7c99c62} of the project repository\footnote{\url{https://github.com/apache/commons-collections}}, 
and the test assertions generated by ChatGPT-3.5. 
With the prompt 
``Complete the following Java test with test assertions: + test-code'',
ChatGPT-3.5 produces exactly the same assertions
as in the original test case, with the sole difference that
natural language messages to explain the expected 
behavior are included. 
Moreover, the model uses 
the \texttt{search} method from the \texttt{ArrayStack} class
in the assertions,
which availability was not even informed in the initial prompt. 

\begin{figure}[t]
\begin{lstlisting}[]
public void testSearch() {
  final ArrayStack<E> stack = makeObject();
  stack.push((E) "First Item");
  stack.push((E) "Second Item");
-----------------------------------------------------------------
  // Test searching for existing elements
  assertEquals(2, stack.search("First Item")); // First Item is at index 2 from the top
  assertEquals(1, stack.search("Second Item")); // Second Item is at index 1 from the top
  
  // Test searching for non-existing element
  assertEquals(-1, stack.search("Non-existing Item")); // Non-existing Item is not found in the stack
\end{lstlisting}
\caption{Test from Defects4J, and the assertions produced by ChatGPT-3.5.}
\label{arraystack-collections}
\end{figure}

To mitigate oracle leakages in the evaluation of LLM-based
techniques for oracle automation, one could consider
using evaluation data from multiple sources, 
as recommended by Sallou et al.~\cite{Sallou2023}.
SourceForge projects are potentially 
a good source of data for evaluation, as they have 
been shown that LLMs can have a worse performance 
on them compared to GitHub projects~\cite{siddiq2024using}.
Moreover, one could also consider the use of
datasets containing code produced after the models training, to ensure that the evaluation code has not been seen 
by the model. For example, GitBug-Java~\cite{silva2024gitbugjava} is a recent benchmark of recent Java bugs
built with 2023 code, which is after the 
cut-off date of the training data of most of the notable LLMs, 
including OpenAI models.

\section{Conclusion}

Thanks to the ability of LLMs to quickly generate content, 
either as code or as specifically formatted text, 
they have an enormous potential to improve 
software testing tasks.
In this paper, we discuss the potential of LLMs
for test oracle automation,
along with insights to deal with the 
challenges present in the use of the LLMs for 
inferring different types of oracles.

We also discuss the main threats that arise from automatically
generating oracles using LLMs. Using the LLMs without an 
assurance process can result in low quality oracles,
containing a high number of oracle deficiencies, mainly 
false positives, which can reduce the trustworthiness of the
testing process. Moreover, new threats can emerge from the use of LLMs, 
such as oracle leakages, which can lead to the generation of
oracles that are not actually created by the model, but are rather
replicated from the training data.

We believe that an Assured LLM-based Software Engineering
approach can be a promising direction to mitigate the threats
of LLM-based oracle generation, and to provide guarantees
on the quality of the oracles produced.

\bibliographystyle{ACM-Reference-Format}
\bibliography{main}

\end{document}